# IRS-assisted UAV Communications: A Comprehensive Review

**Syed Agha Hassnain Mohsan** [1,*], **Yanlong Li** [1,2]

1. Optical Communications Laboratory, Ocean College, Zhejiang University, Zheda Road 1, Zhoushan 316021, China; hassnainagha@zju.edu.cn (S.A.H.M.); lylong@zju.edu.cn (Y.L.)
2. Ministry of Education Key Laboratory of Cognitive Radio and Information Processing, Guilin University of Electronic Technology, Guilin 541004, China

* Correspondence: hassnainagha@zju.edu.cn

**Abstract:**
Intelligent reflecting surface (IRS) can smartly adjust the wavefronts in terms of phase, frequency, amplitude and polarization via passive reflections and without any need of radio frequency (RF) chains. It is envisaged as an emerging technology which can change wireless communication to improve both energy and spectrum efficiencies with low energy consumption and low cost. It can intelligently configure the wireless channels through a massive number of cost effective passive reflecting elements to improve the system performance. Similarly, unmanned aerial vehicle (UAV) communication has gained a viable attention due to flexible deployment, high mobility and ease of integration with several technologies. However, UAV communication is prone to security issues and obstructions in real-time applications. Recently, it is foreseen that UAV and IRS both can integrate together to attain unparalleled capabilities in difficult scenarios. Both technologies can ensure improved performance through proactively altering the wireless propagation using smart signal reflections and maneuver control in three dimensional (3D) space. IRS can be integrated in both aerial and terrene environments to reap the benefits of smart reflections. This study briefly discusses UAV communication, IRS and focuses on IRS-assisted UAC communications. It surveys the existing literature on this emerging research topic and highlights several promising technologies which can be implemented in IRS-assisted UAV communication. This study also presents several application scenarios and open research challenges. This study goes one step further to elaborate research opportunities to design and optimize wireless systems with low energy footprint and at low cost. Finally, we shed some light on future research aspects for IRS-assisted UAV communication.

**1. Intelligent Reflecting Surface (IRS)**

IRS,a promising paradigm, can support high intelligent, controllable and smart wireless communication for future 6G networks [1-2]. It is a large planer surface comprised of massive reflecting elements to introduce desired phase shift to impinging signals. Each element can alter frequency, phase and amplitude of incident signals. A dense deployment of IRS can smartly control the incident signals in order to achieve required signal distributions. Thus, it enables innovative approach to control the wireless channel by attaining a quantum leap gain for secrecy, sustainability, reliability and capacity of wireless communication.

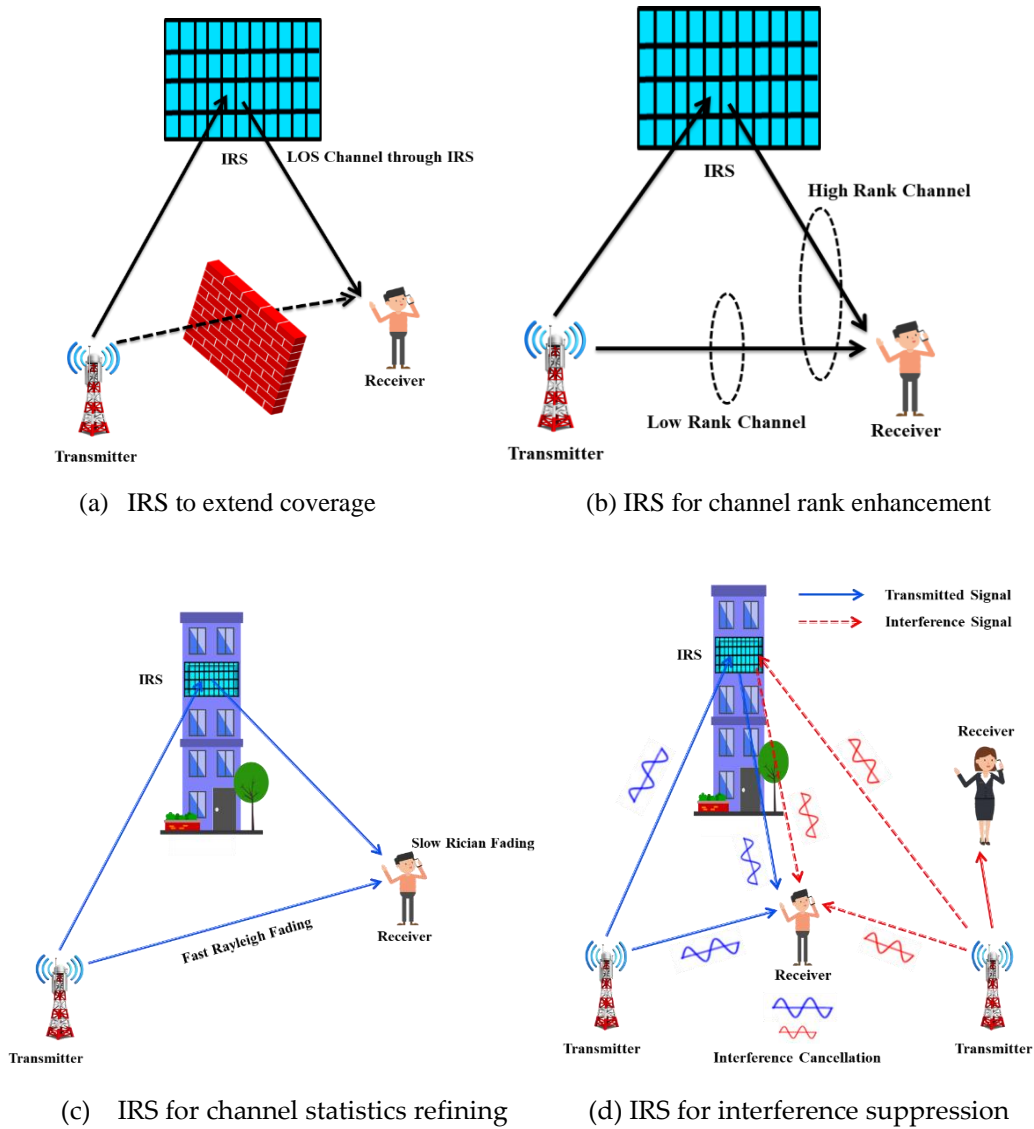

**Figure 1.** Main functionalities of IRS to reconfigure wireless propagation [3]

It can be seen in Figure 1 that IRS supports several functionalities to reconfigure wireless channel including interference suppression, refining channel statistics, coverage extension and channel rank enhancement. In addition, it enables virtual line-of-sight (LoS) link to evade hindrances e.g., buildings and walls between transceivers using smart reflections. IRS supports multiple potential benefits like low energy and hardware cost as IRS elements can alter the signal propagation without any need of transmit power [4]. Apart from this, IRS has low weight, less fabrication cost, low profile and compliant architecture which makes its integration easy in a required object or environment. Similarly, IRS functional mode is full duplex (FD) mode without self-interference and antenna noise amplification. These features make IRS more suitable than half-duplex (HD) relaying systems of low spectral efficiencies. It is also worth mentioning that IRS enables transparent integration, which makes it highly flexible and compatible to current wireless communication technologies e.g., wireless fidelity (WiFi) and free space optics (FSO) All these features place this emerging IRS technology a step ahead to colossally deploy in wireless network for its energy and spectral efficiency enhancement. It can be anticipated from recent studies that IRS will offer novel paradigm shifts from existing massive-MIMO (mMIMO) networks leading to

IRS-aided MIMO networks and heterogeneous wireless networks leading to IRS-assisted hybrid wireless network as presented in Figure 2(a) and 2(b).

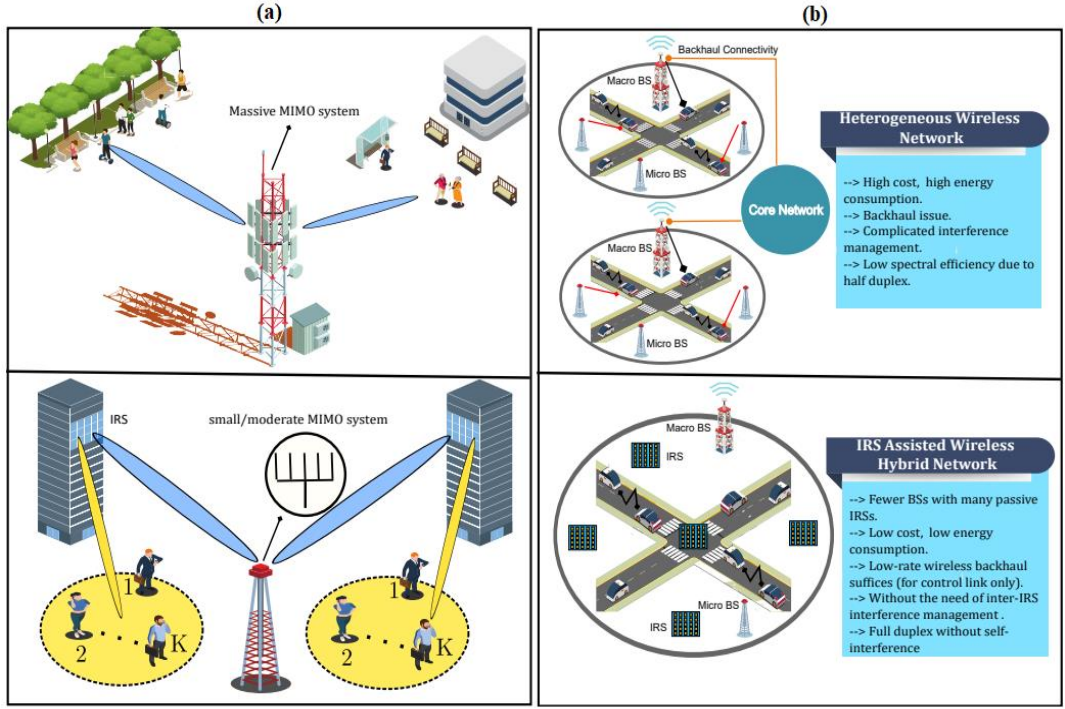

**Figure 2.** IRS paradigm shifts in wireless network, (a) mMIMO and IRS-aided MIMO system, and (b) Heterogeneous network and IRS-aided hybrid wireless networks [3]

IRS-assisted MIMO networks will enable the base station (BS) of a few antennas through a large aperture of IRS to attain three dimensional (3D) passive beamforming via smart reflections [1]. Hence, it can significantly reduce the system hardware cost and energy consumption. IRS deployment in current heterogeneous networks will bring novel hybrid network architecture of both active and passive elements. Moreover, IRS passive reflecting elements are cost-efficient than active elements, so a remarkable cost reduction can be achieved through dense its deployment without any interference control requirements [5]. Major differences between active and passive IRS elements are summarized in Table I.

TABLE I. MAJOR DIFFERENCES BETWEEN ACTIVE AND PASSIVE IRS

| **Active IRS** | **Passive IRS** |
| --- | --- |
| No RF chain | No RF chains |
| High cost | Low cost |
| It can overcome double fading | It suffers from double fading |
| Additional power consumption | Negligible power consumption |
| Additional thermal noise | Negligible thermal noise |

| Signal reflection with amplification | Signal reflection without amplification |

In indoor scenarios, IRS can be integrated on walls, ceiling, room furniture in order to attain enhanced coverage which is crucially important at stadiums, offices, hospitals, shopping malls and airports. In case of outdoor scenarios, IRS might be mounted on building facades, street-lights, advertising boards and traffic signals [6]. These application scenarios will make IRS a revolutionary paradigm towards smart house, smart offices, smart factories, smart hospital, smart factories, smart city and smart oceans. Recently, IRS has received significant interest in future 6G ecosystems [7]. Several research centers and companies have initiated pilot projects on IRS as provided in Table II. In literature, IRS has been used with different notions such as RFocus [8], large intelligent surface/antennas (LISA) [9], reconfigurable intelligent surface (RIS) [10] and smart reflect-array [11].

TABLE II. RESEARCH AND INDUSTRIAL PROJECTS ON IRS

| Year | Research project | Aim/Objective |
|---|---|---|
| 2017 | VisorSurf | It aims to develop a hardware architecture for software-driven metasurface |
| 2019 | ARIADE | It aims to design a metasurface integrated radio and artificial intelligent (AI) technologies |
| 2021 | PathFinder | It aims to develop fundamental algorithms for smart metasurface empowered 2.0 wireless networks |
| **Year** | **Company** | **Aim/Objective** |
| 2018 | NTT DOCOMO and Metawave | To ensure 5G data transmission of 28 GHz-band through metasurfaces reflectarray |
| 2019 | Lumotive and TowerJazz | Experimentally demonstrated the first true solid-state beam steering through liquid crystal metasurface |
| 2019 | Pivotal Commware | Experiments of holographic beamforming technique through software-defined antennas |
| 2020 | Greenerwave | Aims to design physics-inspired algorithmic techniques for controllable binary metasurfaces |
| 2020 | NTT DOCOMO and AGC Inc. | It focuses on developing the first ever prototype of transparent dynamic metasurface for 5G |

*1.1. Prior Research on IRS*

The research on IRS is rapidly expanding around the globe. Research fraternity around the globe is putting efforts to dig into IRS manufacturing, deployment, networking, channel modeling and integration with multiple technologies including backscatter communication, mMIMO, Unmanned Aerial Vehicle (UAV), light fidelity (LiFi) and WiFi. Some recent articles have provided comprehensive review and surveys on IRS [2], [10], [12-13]. Wu et al. [2], highlighted IRS design, deployment and relevant potential challenges. In [10], Di Renzo et al. highlighted the integration or AI and IRS to enable smart radio environments. Mathematical modeling of IRS has been presented in [12]. While Zhao and Yang Liu [13] briefly discussed research progress of IRS. They investigated IRS implementation in several novel scenarios such as secure communication and terminal positioning. In [14], authors proposed algorithm for IRS-aided wireless communication in the perspective of security. Similarly, another recent study [15] investigates hardware and capacity degradation challenges in large intelligent surface (LIS). Several studies have discussed mmWave communication, mMIMO, intelligent mMIMO, large-scale MIMO and holographic mMIMO [16-18]. Some technical studies [19-20] also reported on IRS and MIMO-non-orthogonal multiple access (MIMO-NOMA). An illustrative summary of research contributions on IRS is presented in Table III.

TABLE III. RESEARCH CONTRIBUTIONS ON IRS

| Reference | Research Contributions |
|---|---|
| [21] | In this study, authors briefly discussed a historic overview on wave propagation control for acoustics and optics. They provided an overview of wireless trials at 2.47 GHz in an indoor scenario for spatial EM modulation with an RIS of passive elements. This article also describes the status on RIS hardware architecture, potential challenges and future research aspects for RIS-enabled 6G wireless communications. |
| [22] | In this study, authors have focused on metamaterials as key innovation enablers to address challenges in massive IoT by using RIS. Authors have discussed a B5G scenario with the assistance of RIS to tackle non-line-of-sight (NLOS) paths. |
| [23] | In this article, authors have discussed that RIS offers advantages for mapping and localization in terms of extended physical coverage and enhanced accuracy. This paper aims to facilitate its readers with up-to-date overview of RIS mapping and localization. Authors have discussed applications and challenges associated to RIS. In addition, authors have described several prominent research questions and major avenues to solve these challenges. |
| [24] | In this article, authors have highlighted the impact of hardware imperfection, transceiver antenna misalignment and path attenuation on THz communication. In order to mitigate these challenges, authors used RIS and discussed analytical layout of RIS-assisted THz communication. Moreover, authors proposed a novel approach to optimize the IRS's phase-shifts. |
| [25] | This study surveys the research outcomes in Reconfigurable Intelligent Surfaces (RISs). Authors have emphasized on different aspects for commercial viability of future 6G network deployment. This study also summarizes hardware design considerations, use cases, system models and physical layer optimization methods. |
| [26] | This study reviews IRS-aided non-terrestrial |
| | networks (RANTNs). Authors discussed several properties of aerial RISs (ARISs) and terrestrial RISs (TRISs). They have described different configurations of RANTNs including RIS-aided air-to-air communication, ground-to-ground and ground-to-air and air-to-ground communications. In addition, optimization and performance analysis have been included in this study. This study also highlights open challenges and future research aspects for RANTNs. |
| [27] | This study provides a comprehensive overview on IRS, design aspects, use cases and application scenarios for wireless systems. Authors have studied the performance metrics and provided suggestions about future research directions. |
| [28] | In this article, authors have developed a spectrum learning (SL) assisted RIS framework to smartly exploit the inherent features of RF spectrum for green 6G. The RIS controlled of given architecture has the capability to smartly "think-and-decide" whether to reflect incident signals or not. This study also highlights promising future research directions. |
| [29] | This study presents a tutorial overview of efficient designing of IRS-assisted wireless energy transfer (WET) and wireless information transfer (WIT) from signal processing and communication perspectives. Authors have provided the solutions to overcome potential challenges such as IRS deployment, channel estimation and passive reflection. This study also highlights future research investigations. |
| [30] | This study presents a physical layer security system comprised of an RIS-aided access point, legitimate node and an eavesdropper. The RIS is used to enhance communication and secrecy performance to mitigate eavesdropping attacks. Authors also state that system's performance can be enhance if we can control RIS elements and average signal-to-noise ratios (SNR). |
| [31] | This study presents a tutorial on RIS characteristics from a signal processing aspect by focusing on existing surveys of localization, communication and hardware features. Authors have provides derivations and formulas which are needed to understand RIS systems and how they can be used to enhance sensing, localization and communication. |
| [32] | In this survey, authors have focused on IRS design, potential challenges, applications and future research directions for IRS-empowered 6G wireless communications. Authors have outlined performance analyses and analytical methods to enhance performance of IRS-assisted wireless networks. |
| [33] | This study focuses on research on IRS, working principle and future research direction for smart radio environments empowered by RIS. |

*1.2. An Overview of IRS Technology*

This section discusses IRS technology with respect to its basic architecture and working principle. We have discussed the most common architecture which has been discussed in various existing studies reported on this research topic.

*1.2.1. Basic Architecture of IRS*

The basic architecture of IRS has been extensively discussed in existing studies. Most of the reported studies on IRS propose various designs based on different number of layers. In general, IRS architecture is based on doped semiconductor and electromechanical switches, microelectromechanical systems (MEMS), varactor-tuned resonators and liquid crystals. Although several studies have been reported on this topic, most of these studies focus on three layers (i) a metasurface layer, based on passive conductor; (ii) a control-layer, used to configure phase or amplitude of meta-atom components; and (iii) a gateway layer used to establish communication link between base station and control layer. The key operation of each meta-atom is to configure EM characteristics as a sub-wavelength scatterer. This feature can alter the impinging patterns on IRS into desired EM response [34]. It helps IRS to adjust the wavefronts to attain sensing, collimation, localization, filtering, absorption, beam steering and polarization. Two common ways to ensure steering and focusing are metasurfaces and traditional antenna arrays [35]. The basic architecture of IRS is illustrated in Figure 3.

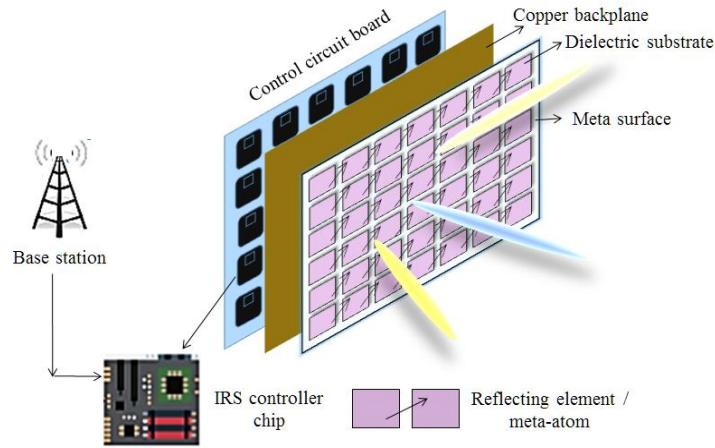

**Figure 3.** Basic hardware architecture of IRS

*1.2.2. Working Principle*

In general, IRS refers to a two dimensional array composed of massive number of passive reflecting elements. The basic principle of IRS is based on changing signals via reflection and controlling their phase shifts. Through this approach, IRS can ensure cost-efficient sustainable performance [36]. In some recent works, the authors have addressed IRS, large intelligent surfaces (LIS) and their role in future wireless networks [37-40]. In [9], Alghamdi et al. have focused on metasurfaces or passive reflecting elements in LIS. These surfaces are based on microelectro-mechanical systems (MEMS) or complementary metal-oxide-semiconductor (CMOS) having the potential to intelligently control phase shifts of impinging signals. Some other switching technologies can be used as well such as liquid crystals or positive intrinsic- negative (PIN) diodes. In the context of PIN, PIN on-off condition can be enabled by an external bias, which can generate two separate levels as indicated in Figure 4. In this case, most of the energy is reflected as PIN diode is on, while most of the energy is absorbed when PIN diode if off.

The EM characteristics metasurfaces are generally associated with meta-atom patterns. Thus, some structure of meta-atoms can fully reflect and others can partially reflect the impinging EM waves. These metasurfaces based on passive reflecting elements are dynamic in nature, thus these surfaces can generate required EM response or can control their conditions by applying an external bias. The scattering elements of these surfaces operate as input/output antennas. Hence, when an impinging EM wave enters from the input antenna, transmission is performed on the basis of switch status, and this EM signals exists from the output antenna with a required reflection.

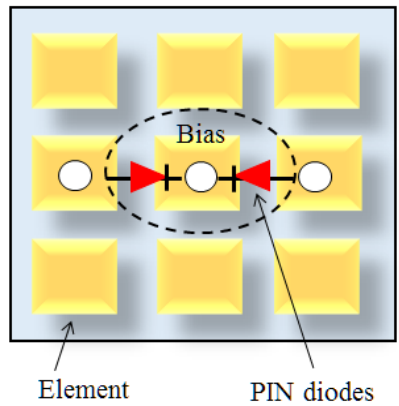

**Figure 4.** EM reflection via PIN diodes

Apart from aforementioned discussion, varactor-tuned resonators have been proposed to control the impinging EM waves as shown in Figure 5. In this scenario, bias voltage is applied to the varactor with the aim to conceive phase shift. The obtained phase shift can be further configured through liquid crystals. Effective dielectric constant can be obtained through altering the DC voltage on liquid crystals. Through this mechanism, the phase shift of incident signals can be adjusted at different locations of the metasurface.

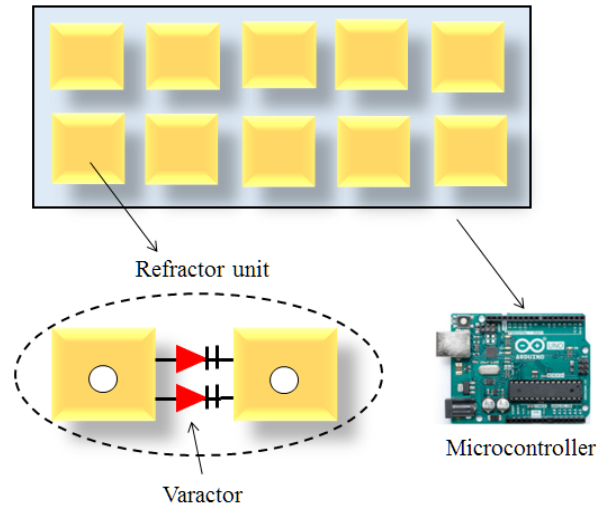

**Figure 5.** Controlling EM reflection via varactor-tuned resonators

## 2. IRS-assisted UAV Communication

In the past few years, we have witnessed significant efforts from both academia and industrial experts to deploy 5G and forecast the roadmap towards future 6G wireless network. 6G is envisaged to support highly consistent network performance as compared to 5G, such as low latency, high reliability, ultra-high data rate, ubiquitous connectivity and high coverage. Particularly, in 6G, the most critical challenge is to achieve high capacity and ultra-reliable wireless communication. Significant research contributions have been made through coding schemes, modulation techniques and space-time-frequency diversity methods to overcome channel fading and shadowing. However, these methods cannot change the wireless propagation, which then arises the innovative idea of controllable and reconfigurable wireless propagation. It can be attained by two methods known as UAV communication and IRS-assisted communication.

Because of cost-effectiveness, high mobility and flexible deployment, UAV has proven its stature to enhance the performance of current cellular networks. UAVs are envisioned to support on-demand and high speed wireless connectivity. UAV communication can be established in disaster stricken areas, where terrene networks are damaged. For instance, UAVs have been successfully used to monitor disaster in Henan, province, China in 2021. In particular, by exploiting high mobile and controllable UAVs, the UAV trajectory can be smartly controlled to achieve intended communication. UAVs offer better wireless channels by establishing LoS air-ground communication links. Enhanced quality of communication link may be obtained by smartly adopting the trajectory or hovering position to shorten the distance of LoS links. Similarly, the high mobility of UAVs also supports enhanced performance in air-ground networks. Research works have been carried out to dynamically optimize the UAV trajectory according to surrounds and users to meet the QoS requirements. To ensure higher capacity, UAVs might be incorporated with the emerging technologies at the physical layer, including millimeter wave (mmWave) and non-orthogonal multiple access (NOMA) etc. Several use cases of UAVs have been investigated to enable stable performance considering reliability, security, capacity and throughput. Furthermore, UAVs can operate as friendly jammers to enhance network security [41]. In literature, several studies have been reported on UAV communication [42-44]. The research and development initiatives are still in progress to discover diverse breakthroughs to leveraging the UAV communication performance. Still there are several areas which are yet to be explored for real-time implementation of UAV communication. For instance, certain limitations on battery endurance, flight time and UAV payload still make it hard to integrate with promising technologies such as MIMO to ensure flexible beamforming to overcome high path loss. UAVs impose critical limits due to their stringent size, weight, and power (SWAP) constraints. Despite the appealing benefits of UAV communication such as LoS link due to its elevated altitude, still the air-to-ground channel can face blockage due to complicated terrain and surroundings such as high rise buildings. Particularly, besides transceiver power consumption, UAVs consume additional energy to achieve high mobility and remain airborne. Furthermore, the privacy and security of legitimate network entities might be at a risk due to eavesdropping. These challenges cause a significant decrease in performance of UAV communication.

IRS is an emerging technology which can overcome aforementioned challenges. IRS is envisaged to revolutionize the future wireless networks as it has the capability to improve the communication capacity and coverage. In terrestrial networks, IRS has been used to improve physical layer security, energy efficiency and capacity. The phase, refraction, reflection and absorption of the IRS passive reflecting elements can configure the incident EM wave in the intended direction. The amplitude and phase of the reflected wave enhances the effective channel gain. Compared with current techniques, such as backscatter communication and active relaying system, IRS can be easily integrated at various locations including indoor walls, vehicles and buildings with minimal effort and cost. IRS is more energy efficient as it can passively reflect the incident signals without requiring RF chains. Additionally, it offers the benefit of its compatibility with existing radio technology. IRS supports wide range of signal reflection and more flexible deployment when it is installed on a mobile UAV rather than a fixed building. It offers both half-duplex and full-duplex communication over wide frequency and bandwidth along with various design options. Thanks to its conformal geometry, IRS installed on facades of buildings can establish LoS paths with the UAV due to shorter distance and high altitudes. It supports circumvent environmental obstructions by offering LoS link between blocked user and UAV through reflected link by IRS. Similarly, IRS-enabled UAV communication can also reduce the propulsion energy utility for UAVs and significantly minimizes the access delay. As shown in Figure 6, deploying IRS with UAV communication has great ability to overcome the increasing complexity of the wireless channel. IRS can be used to expand UAV coverage and can achieve several QoS requirements of users. IRS can be integrated with UAVs to attain a desired propagation scenario and enhance the communication performance of intended users. Meantime, IRS can neglect the unintended signals to overcome interference and avoid adverse eavesdropping. IRS-aided UAV system can enhance the channel quality in urban scenarios as presented in [45]. The researchers derived a closed-form approach for the phase shifter as function of UAV path and present a significant enhancement in achievable rates over time. Moreover, authors in [46] demonstrate how IRS-assisted UAV communication can enhance cellular applications. The IRS-UAV concept has a viable reduced power consumption as compared to active ARs, which can be further enhanced by a jointly optimizing the resource allocation and UAV trajectory [47]. Authors in [48] consider a broader aspect to discuss the full potential of IRS-assisted UAV framework. In this study, they discussed several use cases, challenges and future research opportunities for IRS-assisted UAV communication in the context of future cellular networks. They discussed related works in the domain of physical layer security, spectrum sharing, giant-site access and enhanced coverage. In [49], authors discuss the deployment of IRS on ground facilities. By considering the transmission from UAV to ground users, a ground-deployed IRS is used to guide the signals towards desired users, thereby achieving enhanced performance and reducing path loss even in the absence of beamforming at the UAV. This study also discusses UAV airborne IRS-enhanced communication. There are several research efforts dedicated to IRS-assisted UAV communication [50-57]. In [58], authors combined IRS and UAV to significantly enhance the performance of air-ground network by taking both aerial and terrestrial IRSs into account. They considered two case studies to highlight the advantages of combining IRS and UAV. In the first scenario, the IRS is incorporated on a UAV to improve the communication performance from the base station to the user. While in other case, the IRS is used to support UAV to the ground transmission. In several existing studies, a terrestrial IRS is installed at a fixed position which supports only nearby users. Similarly, in complex urban scenarios, the signal might be reflected by multiple IRSs to bypass any blockage. Consequently, it leads to high path loss [59-60]. To overcome this challenge, a possible solution is to mount an IRS onto the UAV in terrene communication. The UAV can be tethered with ground-based power supply or moving vehicle to ensure stable control and reliable power supply. In Table IV, we have summarized some existing studies on IRS-assisted UAV communication.

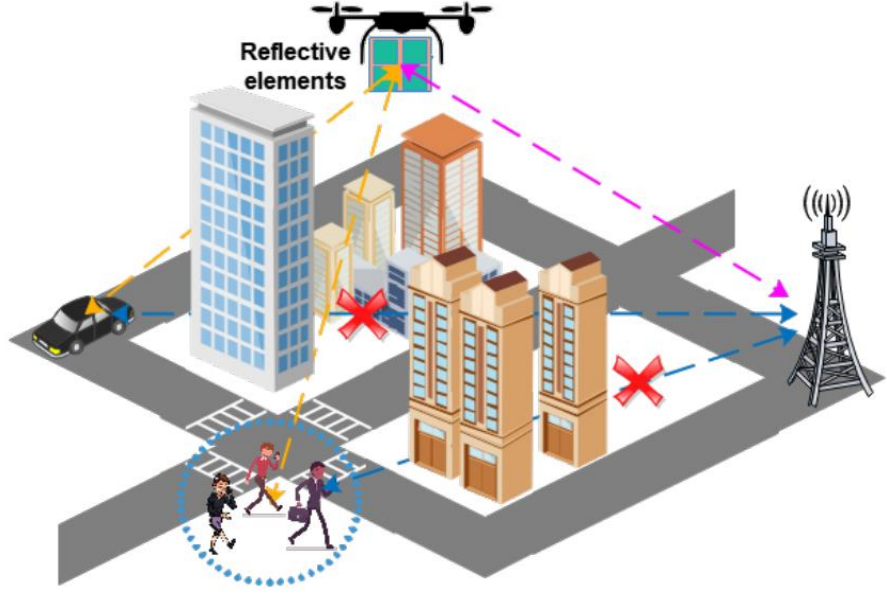

**Figure 6.** IRS-assisted UAV communication [48]

**Table IV.** Existing works combining UAV and IRS

| Reference | Aim | Optimization variable | UAV mobility | IRS installation |
|---|---|---|---|---|
| [45] | To maximize the average achievable rate | UAV trajectory, IRS passive beamforming, | Mobile | At the building |
| [53] | To maximize the transmission capacity | Reflection and location parameters of IRS-UAV | Static | At the UAV |
| [54] | To maximize the rate of strong user | Location, phase shift and beamforming of IRS-UAV | Static | At the UAV |
| [55] | To maximize the secrecy rate | IRS phase shift, power control of UAV, trajectory | Mobile | At the building |
| [56] | To maximize the received power | IRS passive beamforming, beamforming and trajectory of UAV | Mobile | At the building |
| [57] | To maximize the IRS data transmission | IRS scheduling, IRS phase shift, UAV trajectory | Mobile | At the building |

*2.1. Advantages of IRS-assisted UAV Communication*

The key potentials of IRS-assisted UAV communication can be listed as follows:

• In case of blockage or absence of direct LOS links, Virtual LoS links might be designed between ground users and UAV through IRS in air-ground networks. Hence, the communication quality can be significantly improved, which ultimately extends the network coverage.

• Being mounted on a UAV, the IRS location can be dynamically and smartly controlled as well as UAV's mobility. The mobility of UAV enables a new degree of freedom for IRS design as compared to terrene IRS installed on fixed places.

• When installed on a UAV, it is highly possible that IRS can establish an LoS path with both transmitting and receiving components. In comparison with the terrene IRS which supports the communication between two nodes positioned in half of the space, installing IRS to a UAV can expand the wireless coverage and attain the full-angel reflection.

• Both UAV and IRS can also operate in the terahertz (THz) or mmWave band, enabling an ultra-wide bandwidth to satisfy the requirements for high data rate. Nonetheless, the THz or mmWave links are severely degraded by high path loss and blockages, which can be efficiently compensated through the mobility of UAV and reconfigurability of IRS.

• Another motivation behind IRS-assisted UAV communication is to enhance achievable data rate and signal-to-noise-ratio (SNR). In [61], authors considered ground-based IRS to achieve ground-air UAV communication improvement. They considered different aspects like number of IRS reflecting elements, IRS location and UAV hovering. Authors formulated an optimization approach to improve the received signal power at the UAV by the optimization of IRS location and phase shift of reflection elements. Similarly, airborne-IRS can be integrated with other promising technologies such as mmWave, FSO and NOMA to enhance the communication data rate.

• In UAV communication, the link suffers from LoS blockage. The UAV is always observable to RIS, ground users and potential eavesdroppers. The communication link reliability is degraded by eavesdropping channel interference. To overcome this issue, IRS can be used to mitigate external interference. An overview of IRS-aided UAV communication in the presence of a potential eavesdropper is shown in Figure 7. In such case, the eavesdropper can intercept and interfere in wireless propagation environment, which may lead to information theft and security concerns. Through passive beamforming, IRS enables a directional beam to decrease the received SNR at the eavesdropper and strengthens the reflected signals for the intended users. Thus, it can improve the physical layer security and reliability for UAV communication. As fixed deployment is not suitable for mobile users, airborne IRS can support a substantial performance enhancement for the legitimate users after the optimization of the UAV trajectory or location.

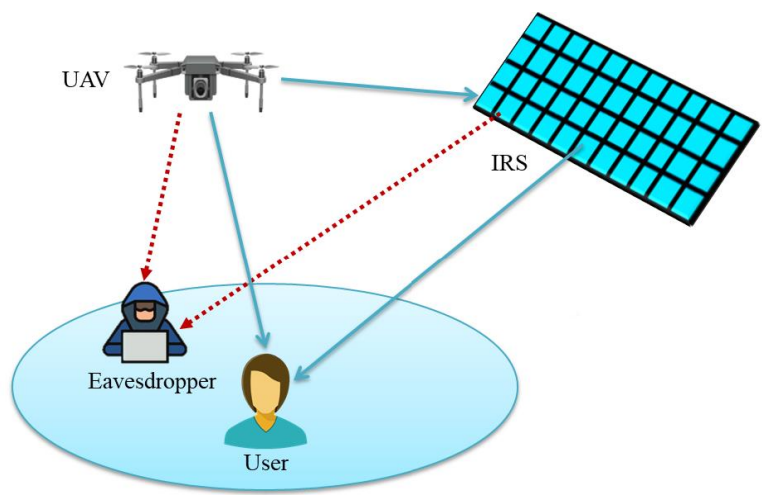

**Figure 7.** IRS-assisted UAV communication in the presence of eavesdropper

• The energy consumption of UAV is a critical challenge in in UAV communication due to limited

batteries and payload. As IRS has the capability to enhance achievable rate and SNR, researchers have been working on IRS-assisted UAV communication to overcome the energy consumption issue. In [62-63], IRS phase shift, UAV trajectory and power allocation are collectively optimized to reduce the average energy consumption. In [63], authors used deep reinforcement learning (DRL) to optimize time-varying channels in a dynamic environment, thus leading to enhanced network performance and reduced energy consumption. In [64], authors proposed an extended probabilistic LOS model and analyzed the benefits of IRS deployment in UAV communication. The communication rate at the users can be enhanced by using IRS along with reducing the energy consumption for UAV. In [65], authors discussed a communication system with IRS and cooperative UAV. A directional antenna is mounted on UAV pointing towards IRS to minimize the energy consumption.

Despite the aforementioned benefits, IRS-assisted UAV communication suffers from various challenges such as channel estimation, location of IRS, orientation of IRS and controllability or endurance of UAV. For the practical implementation of IRS-assisted UAV system, the UAV's and IRS's practical constraints should be kept into consideration along with transmission design.

## 3. Emerging technologies

Recent works show that IRS-assisted UAV communication can support tuned channel gains, improved QoS, extended coverage, high reliability, secure transmission and reduced energy consumption. In this section, we have discussed the integration of IRS-assisted UAVs with different emerging technologies such as machine learning (ML), free space optical (FSO) communication, visible light communication (VLC) and non-orthogonal multiple access (NOMA).

*3.1. Machine Learning*

The IRS-assisted UAV communication has proven its stature is several applications. Novel breakthrough novelties can be conceived through the integration of artificial intelligence (AI) and machine learning (ML). These promising technologies can support IRS-assisted UAV communication in the perspective of network performance, security, reliability and QoS. These technologies can ensure potential characteristics such as data extraction, smart decision-making and optimal optimization. ML algorithms can be implemented to enhance spectral efficiency, UAV tracking performance, embedded optimization, IRS channel estimation and managing several tradeoffs. In addition, ML approaches can be utilized to smartly control IRS's phase shift and UAV's location in order to obtain desired communication path. Similarly, reliable channel characteristics and precise tracking can be achieved through ML tools. Furthermore, deep neural networks (DNNs) and convolutional neural networks (CNNs) can be implemented to reduce the computational time and complexity of these tools.

*3.2. Free Space Optical (FSO) Communication*

Optical wireless communication such as FSO communication has emerged as a promising technology to meet the rapidly expanding needs of 5G and B5G technology [66]. FSO systems offers huge bandwidth spectrum to feature high data rate services such as wireless backhauling, tactile internet and video surveillance. This technology ensures unique advantages such as low cost components, simple system and easy deployment. However, FSO systems are vulnerable to several shortcomings in the most critical ones are misalignment or LOS constraint. In addition, FSO systems are prone to atmospheric turbulence, heavy snowfall and fog. To overcome these issues, several methods have been suggested in literature including adaptive optics, diversity techniques, beam steering and aperture averaging. Some studies have addressed these critical concerns but these studies are in sparse [67]. IRS-UAV aided FSO communication has a viable impact to design a flexible communication and energy efficient system. The authors in [67] generalized pointing error loss using Hoyt distribution and atmospheric turbulence through Gamma-Gamma distribution in order to assess the performance of proposed system. The proposed system model was validated using Monte Carlo simulation.

*3.3. Visible Light Communication*

Another emerging candidate which will support promising features of future 6G wireless networks is VLC as it has unlicensed spectrum, ultra-high data rates and low implementation cost. However it has some critical drawbacks including LoS requirements, misalignment, pointing error, and limited coverage. In such scenario, IRS-assisted UAV system can mitigate these shortcomings and enhance VLC system performance. The IRS used to assist in VLC system to adjust the incident optical beam differs from standard IRS. Crystal liquid or meta-lens based IRS can be considered in such systems. Such types or IRS can control the incident optical beams through refractive index or dynamic artificial muscles [48]. UAV's mobility can offer accurate and fast alignment between the optical transmitter e.g., LED or LD and the IRS to establish LoS path. The phase shift of IRS can be adjusted to enhance achievable data rate and extend coverage range. The IRS reflecting elements can adjust the incident light shape through adapting the channel configuration. It is possible to ensure this phenomenon by smartly controlling the IRS thickness and refractive index.

*3.4. Non-orthogonal Multiple Access (NOMA)*

In UAV-empowered communication, UAVs are required to provide services to a massive number of users with stringent requirements. To satisfy these requirements, advanced multiple techniques are important. In particular, non-orthogonal multiple access (NOMA) has emerged as a promising technology to integrate into B5G networks due to the superior performance of user fairness guarantee, massive connectivity and enhanced spectral efficiency. By invoking successive interference cancellation (SIC) and superposition coding (SC) techniques, NOMA allows multiple users to share the same communication resources including space, time, frequency and spreading code. The integration of NOMA in IRS-assisted UAV communication is conceived to be an attractive strategy due to below aspects:

• Firstly, compared with traditional orthogonal multiple access (OMA), NOMA offers efficient resource allocation and flexibility for IRS-assisted UAV communication. Hence, diversified communication demands of users can be satisfied along with enhanced spectrum efficiency.

• Secondly, in traditional NOMA transmission, SIC decoding is performed by separating users as strong and weak user according to channel conditions. The channel condition of users can be degraded or enhanced through adjusting IRS phase shift and UAV's mobility, thus ensuring a "smart" NOMA operation can be achieved. In future, efficient algorithms must be developed to fully reap the potentials of NOMA and IRS for UAV-empowered communication. Several studied have reported the smart integration of these technologies such as [68] studies how to attain a flexible and high coverage performance of a cellular network that empowers UAVs for NOMA transmission to serve multiple users. This study analyzes the downlink coverage of IRS-UAV aided NOMA transmission considering two users. The coverage analyses include the optimal transmit power allocation between UAVs and NOMA users to support flexible and ubiquitous NOMA transmission. In [69], authors investigate IRS enhanced multi-UAV NOMA networks. A novel transmission approach is introduced, where multi-UAVs-enabled BSs employ NOMA to multiple users using an IRS. The transmit power and 3D position of UAVs, IRS phase shift and NOMA decoding are jointly optimized to maximize the sum rate of network. The results indicate that proposed approach can significantly improve the achieved sum rate. Similarly, NOMA performance gain is enhanced by optimizing the UAV's position as it supports a flexible decoding order design. Authors in [70] consider that the UAVs carrying IRSs can play the role of aerial reflectors to reflect the users signals to the required destination, and use the PD-NOMA transmission in the uplink. In [54], IRS-UAV enabled cooperative downlink transmission is considered where the UAV carrying IRS operates as a node between multiple users and BS. This study also considers NOMA to further enhance the spectral efficiency of the system. This study aims to enhance the rate of strong user while guaranteeing the target area of the weak user through optimized position of UAV. In [71], authors investigate the performance of IRS-aided NOMA system, where IRS is installed on a UAV to support the transmission from a BS. The BS uses the UAV as a relay to serve multiple ground users. Authors highlight the advantages of IRS-assisted UAV relaying system and discuss the benefits of combining these three technologies to further enhance the system performance.

## 4. Applications of IRS-assisted UAV communication

There are several use cases of IRS-assisted UAVs in networking and communication. In this section, we discuss the impact of this coexistence in terms of coverage, capacity, PLS, IoUT, spectrum sharing, vehicular communication, smart city and ground-based wireless networks.

### *4.1. IRS-assisted UAV Communications for Extended Coverage*

The aerial user equipment (AUE), aerial relay (AR) or aerial base station (ABS) are envisaged as key enablers to support adaptive and dynamic coverage in future wireless networks. However, the active aerial communication incurs significant energy overhead. IRS has emerged as a promising alternative. A smart deployment mechanism is spherical type IRS which can support wireless propagation control at desired depths, heights and locations. This sphere type IRS will be an innovative design with the capability to reflect incident signals in all directions. In such condition, various geometric bodies such as UAVs or AUVs can be integrated to support aerial or underwater applications. It can reflect signal in multiple directions to cover dead zones rather than a fixed direction. It will support an enhanced spectral efficiency with 360 degrees coverage. In spherical IRS, the total number of reflective components used in signal steering in a specific direction is less than planar IRS. Even though spherical IRS geometry possesses a larger size as compared to planar IRS, howbeit, it also supports various potentials to overcome this drawback such as independent operability of IRS to its orientation. This feature has unique benefits in aerial scenario because of potential rotational mobility of IRS in case of IRS-assisted UAVs. A UAV-carried IRS can offer smart reflections at suitable heights to enhance coverage. Furthermore, optimizing UAV's trajectories and IRS's phase shift, the network coverage in any desired direction can be significantly enhanced. An overview of IRS-assisted UAV scenario is presented in Figure 8.

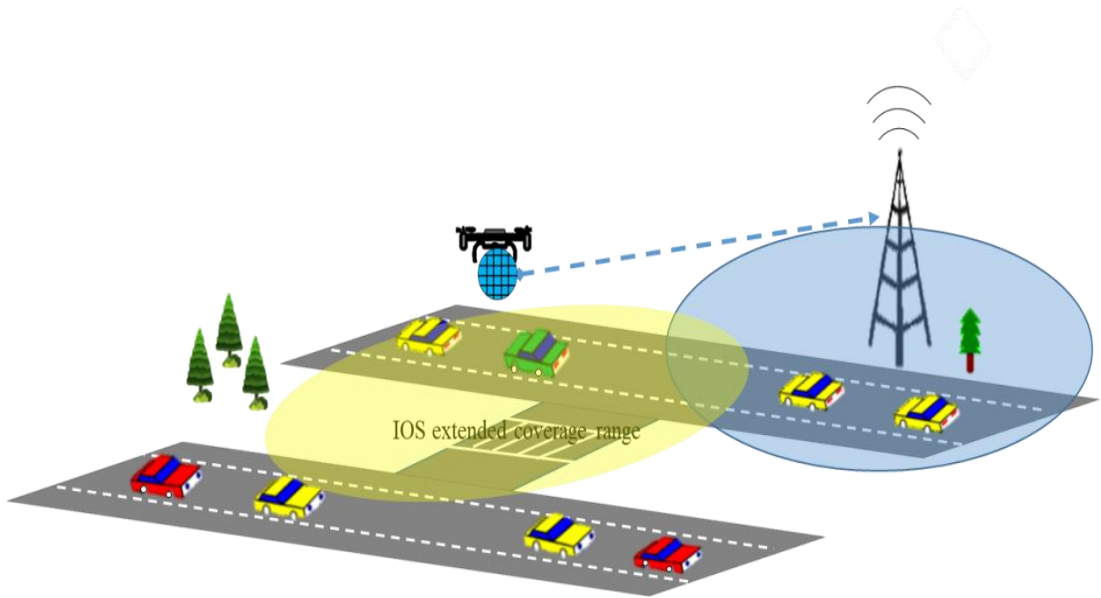

**Figure 8.** IRS-assisted UAV communication for extended coverage

*4.2. IRS-assisted UAV Communications for Increased Capacity*

The achievable rate for both downlink and uplink between UEs and BSs can be maximized through proper positioning of AR. Different optimization problems can be formulated for the UAV orientation, transmit power, trajectory and position. The IRS-assisted UAV is a promising alternative candidate to its counterparts to enhance the capacity and throughput. Similarly, the operational mode of IRS is full duplex (FD) without any self-interference and antenna noise amplification. These features make IRS more suitable than half-duplex (HD) relaying systems of low spectral efficiencies. A proper optimization of IRS technology places it a step ahead to colossally deploy in wireless network for its energy and spectral efficiency enhancement. Generally, the potential to optimize the IRS geometry is restricted by limited knowledge about the associated channels. Without sufficient information about the underlying propagation environment, the IRS cannot efficiently enhance the system performance. To enable the joint beamforming, the accurate information about the channel state information (CSI) is needed. The acquisition of timely and precise CSI has a major contribution in IRS-assisted wireless networks. Considering indirect channels, it is difficult to obtain the CSI where IRS without transceiver chains is employed. As a result the particular signal on IRS cannot be accessed. Most of the times, IRS is configured containing higher number of metasurfaces, resulting into high channel dimension and overhead on the CSI acquisition. Current studies on IRS usually suppose perfect CSI at the BS, IRS controller and users. However, it cannot be taken perfect and requires a very low overhead. Machine learning and deep learning algorithms can be used as a mitigative approach to solve this CSI challenge. Moreover, the IRS can assist UAV to provide rich scattering of LOS paths for multiple ground users. Consequently, considering both UAV's LoS capabilities and IRS's characteristics, the network capacity can be improved. Scalability can be further enhanced through multi-UAVS incorporated with static IRS.

*4.3. IRS-assisted UAV Communications for Spectrum Sharing*

Considering diagonal channel matrix, IRS can offer significant reduction in interference when devices carry our transmission using same frequency slot. This potential feature make IRS suitable candidate for spectrum sharing. The conventional spectrum sharing methods usually consider cognitive radios and require a reliable and effective spectrum sensing strategy to reduce the interference to primary users. Howbeit, it is achieved at the expense of energy and compromise over reliability in complex channel scenarios. In case of UAVs, the energy consumption is of vital significance to carry out extended flight for long-term mission. Thus, IRS-assisted UAV can further enhance the network capacity using spectrum sharing. In [72], authors have empirically presented the benefits of IRS supporting spectrum sharing in indoor environments. In this work, users interference in controlled through IRS phase shift and capacity is increased using multiple access in spectrum sharing strategy. In another recent study [73], authors proposed IRS-assisted spectrum sharing method to enhance secondary users capacity while maintaining QoS for primary users via precise optimization of IRS's phase shift to attain channel diagonalization. Similarly, these works can be further modified through IRS-assisted UAV where UAV characteristics will play a significant contribution to optimize the wireless network performance in real-time application as presented in Figure 9. By installing IRS on a UAV, it is worthy to notice the impact of parameters such as latitude, longitude and altitude on IRS's phase shift to improve system capacity.

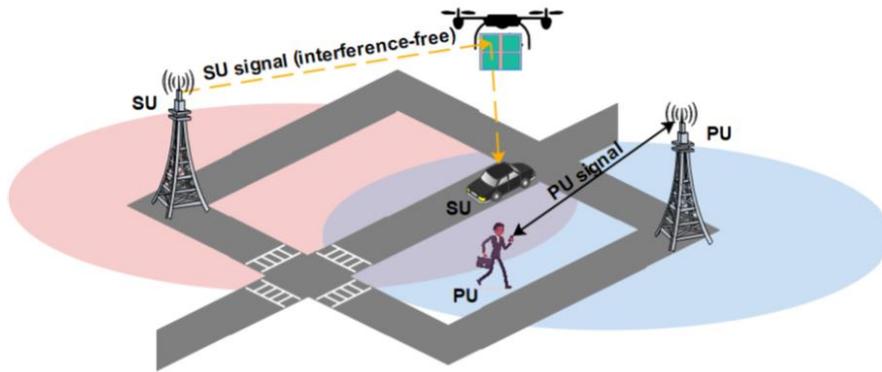

**Figure 9.** IRS-assisted UAV for spectrum sharing [48]

*4.4. IRS-assisted UAV-Ground Wireless Networks*

Even though UAVs can develop LoS links in most application scenarios, but they still suffer from blockages due to obstacles. Additionally, UAV-based ground communication faces severe interference and may be exposed to eavesdroppers. To overcome these challenges, IRS may be deployed to support the intended signals at the desirable users and mitigate obstacles. Furthermore, IRS has the capability to deteriorate the unintended signals at the undesirable users to avoid data leakage and the interference. Generally, there exists a tradeoff to collaboratively adjust the phase shifts to attain these two objectives. In [58], authors have discussed IRS-aided UAV wireless communications to suppress the unintended signals and improve the intended signals respectively. Figure 10 shows an IRS-aided UAV communication in the presence of an eavesdropper to verify the capabilities of IRS by enhancing the security for UAV-based ground networks.

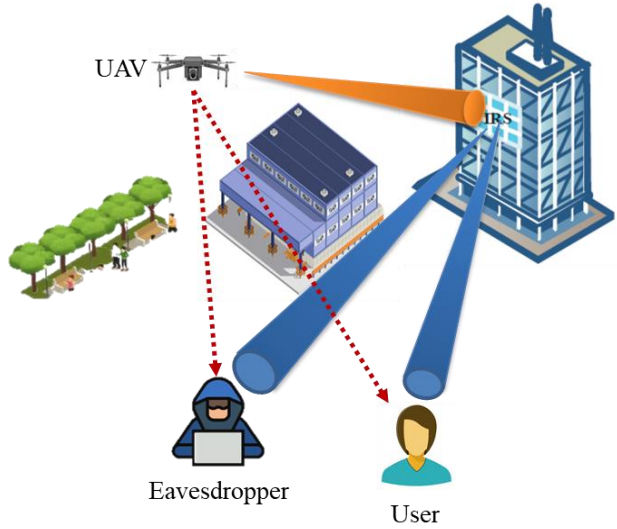

**Figure 10.** An IRS-assisted UAV system in the presence of eavesdropper [58]

*4.5. IRS Enhanced PLS for UAV Communications*

UAVs have the potential to improve the PLS for terrene cellular networks. It can be achieved through dominant LoS path established between ground and aerial nodes. PLS can leverage the characteristics of wireless transmission to enhance the received information at the legitimate user than eavesdropper. There are different methods to obtain effective PLS, such as jamming, transmission power control, signal processing and coding techniques [74]. PLS can ensure secure and efficient defense, authentication and key generation against potential threats. A key benefit of PLS for security is low computation cost and packet size. Therefore, PLS is suitable for energy constraint objects like IoT terminals and UAVs. UAV can operate as an AR between legitimate nodes to weaken the transmission for eavesdropper. Moreover, UAVs may be used as friendly jammers to transmit artificial noise toward eavesdropper to protect legitimate users. Similarly, IRS can play a major role to improve the data confidentiality of UAV communication by offering an improved secrecy rate. IRS scattering elements can ensure desired phase shifts to design a destructive reflective signal to reduce the received SNR and restrict the attackers. The RIS-assisted UAV can smartly tackle the potential attackers. Besides security, IRS-aided UAVs can support energy efficient and reliable communication, thus enhancing the battery life of UAVs. In Table V, we have summarized recent research contribution on IRS enhanced PLS for UAV communications.

**Table V.** Existing works on IRS enabled PLS for UAV communication

| PLS aspect | UAV aspect | IRS aspect | Scenario | Results |
| --- | --- | --- | --- | --- |
| Maximize secrecy rate [75] | Power control, trajectory control | Phase shift control | Single eavesdropper, UAV to single receiver | Enhanced secrecy rate |
| Maximize secrecy rate [76] | Position design, Beamforming | Position design, beamforming | Single eavesdropper, | Enhanced secrecy rate |

|  |  | design | design | UAV BS to receiver |  |
| --- | --- | --- | --- | --- | --- |
| Maximize secrecy rate [77] | Position design, power control | Phase shift control | Single eavesdropper, UAV to ground user | Enhanced secrecy rate |
| Maximize secure EE [78] | Trajectory design, power control | User association, phase shift control | Single eavesdropper, IRS equipped UAV, BS to users | Enhanced secure EE |
| Maximize secrecy rate [79] | Trajectory design | Beamforming design | Single eavesdropper, UAV to ground user | Enhanced secrecy rate |
| Maximize secrecy rate [80] | Position design | Phase shift control | Single eavesdropper, BS to users | Enhanced secrecy rate |

*4.6. IRS-UAV enabled Smart City*

The concept of smart city has emerged from the digital transformation of several ecosystems and harmonization of digital technologies at a city level. To feature a smart city, future wireless technologies should be smartly developed for careful utilization of city resources. This idea can be useful in advance monitoring capabilities of societal processes, advance digitalization of urban scenarios and it can enhance accessibility to public services. In the foreseeable future, smart city components are envisaged to be more controllable and unified with potentials of self-management. Lately, the concept of "smart radio environment" has been introduced for better performance of wireless networks [33]. This concept is linked with the recent advancements in design of intelligent reflecting surfaces (IRSs). Therefore, IRSs are envisioned to bring the required flexibility to develop "smart radio environment" or "smart communication environment" by supporting cost and energy efficient transmission. By deploying IRSs, the ecosystem of the smart city will be more adaptable and controllable, which would subsequently enable easy installation of future wireless communication in urban scenarios and boost the interconnectivity among publication and private applications. In [81], S Kisseleff et al. proposed the idea of RIS integrated smart cities by discussing some new applications, use cases and potential advantages. They considered the fundamental architecture of a smart city model with entities like sensors/actuators, communication platforms, autonomous vehicles and services/applications. They identified the relevant application scenarios as smart communication, smart billboards, smart hospitals, smart factories, smart buildings and smart homes. Authors pointed out most promising opportunities and research directions. In this study, authors also focused on key-enabling aspects of IRS-UAV aided smart city requiring substantial research contributions in RIS control, distributed operation, precoding for large multiuser networks and pilot decontamination. IRS-UAV empowered smart city will enable enhanced QoS, better resource allocation and enhanced security. Figure 11 illustrates IRS-UAV empowered smart city where IRS can be deployed at various objects including stationary (buildings) and mobile (vehicles and UAVs). A key benefit of integration of IRS and UAV in smart cities is to reduce the energy expenditure for wireless systems. It will also mitigate the ecological and health concerns due to proliferation of wireless technologies, especially by relieving human exposure to harmful RF radiations and operating in higher frequencies like THz and mmWave communications.

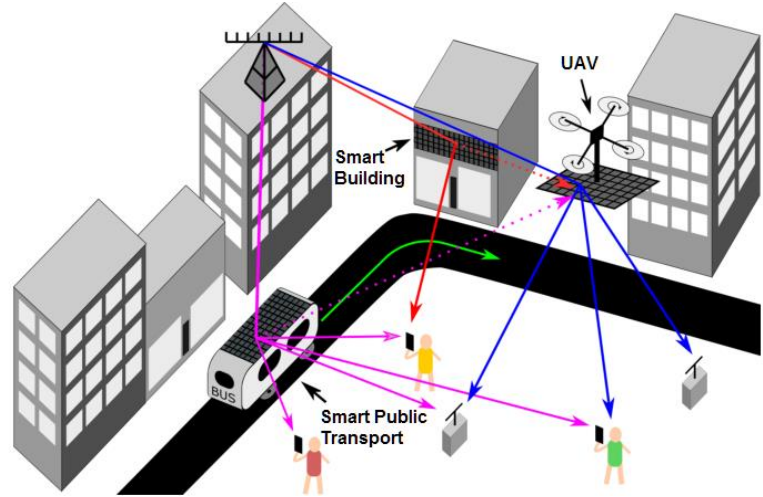

**Figure 11.** An overview of IRS-empowered UAV in smart cities [81]

*4.7. IRS-assisted UAV vehicular communication*

Vehicular communication networks are extensively explored to support cutting-edge smart services including self-driving vehicles and intelligent transportation systems (ITS). Several applications of vehicular communications have been reported such as a wide variety of entertainment, safety, control, autonomous and comfort-oriented applications. For instance, laser imaging detection and ranging (LIDAR) technology has been integrated in vehicular communication networks to achieve data rates in Gb/s. Researchers have started investigating mmWave of 10 GHz to 300 GHz frequency range for vehicular communication [82]. Millimeter wave communication focuses on the limited bandwidth issue in cellular networks. Howbeit, it severely degrades due to high path loss, obstacles which block communication links and thus leading to low transmission distance than low-frequency bands. In addition, beam tracking and transmission outage are yet to be explored in mmWave communication. Similarly, extreme narrow beam of mmWave also arises the critical concern of legitimate user blockage and secrecy information leakage to eavesdropper because of beam misalignment.. To tackle these shortcomings, IRS has been proposed as an innovative and cost-effective paradigm to strengthen signal quality and extend coverage through passive reflecting elements. To reduce blockage, IRS-assisted UAV system supports reliable communication links for mmWave communication. AU Makarfi et al. [83] introduced RIS-empowered vehicular networks to improve PLS. J. Wang et al. [84] also reported outage analysis for IRS-aided vehicular communication. It is envisaged that IRS-UAV system can support traffic monitoring, traffic accident prediction and avoid traffic congestion. An overview of IRS-assisted UAV system for vehicular communication network is illustrated in Figure 12.

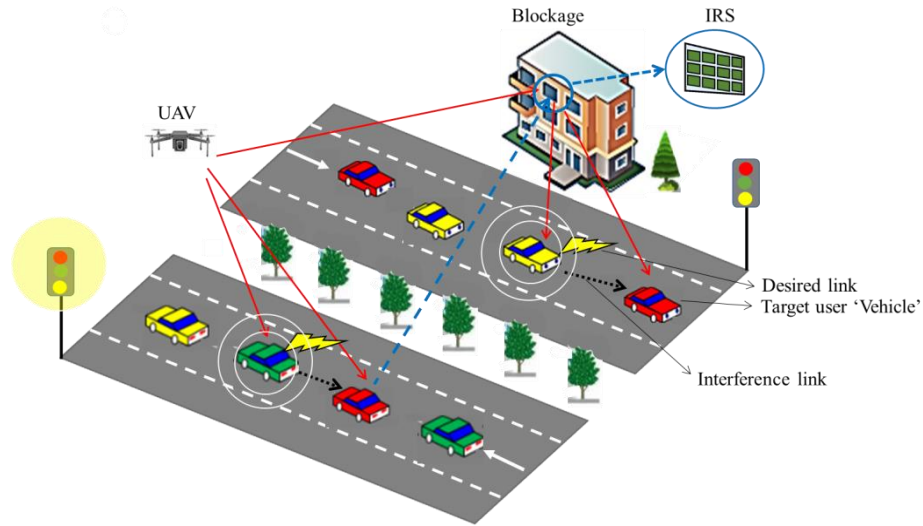

**Figure 12.** An overview of IRS-UAV assisted vehicular communication network

*4.8. IRS-assisted UAV communication in IoUT*

In underwater environment, EM wave or optical communication is limited to a short distance while larger transmission range can be achieved by using acoustic signal in the context of Internet of Underwater Things (IoUT). If case of acoustics communication, it is highly vulnerable to scattering at rough surfaces, suspended particles, water streams and water species. It introduces high path loss, so it can withstand only narrow signal bandwidths. As a result, it decreases the effective data rate. IRSs have been introduced to tackle these issues. Figure 13 outlines the integration of UAV and IRS in underwater environment. IRS can support beam steering in the intended direction which can decrease the multipath effect. Even though it is hard to completely vanish the multipath, however it is possible to optimize IRS scattering elements in order to decrease the frequency selectivity and thus offering enhanced signal bandwidth. IRS can be installed in underwater environment in different geometries based on location and application scenario. It can float below water surface, it can be mounted on an AUV or placed on the shore. AUVs can be preferred to carry IRSs. IRS can be installed on AUVs in order to reduce any vulnerability and preserve control on the smart environment. This coexistence of AUVs and IRSs can support beam steering through IRS phase shift and dynamic mobility of these autonomous underwater vehicles.

Another installation strategy is floating spherical IRS. In this scenario, a wired mechanism like cable can be used to attach IRS to the ground as illustrated in Figure 13. This strategy helps to deploy IRS at desired positions and depths. This floating sphere type IRS can steer incident signals in all directions. In spherical IRS, the number of reflective elements used to steer the signal is less than planar IRS. Similarly, UAVs above the water surface can be used to collect data from buoys or ships and forward to remote monitoring center. IRS-assisted UAV communication requires further research contributions to fully deploy this strategy to support IoUT features in order to support underwater communication, monitoring, data collection, disaster prediction, secure harbor and ocean transportation etc.

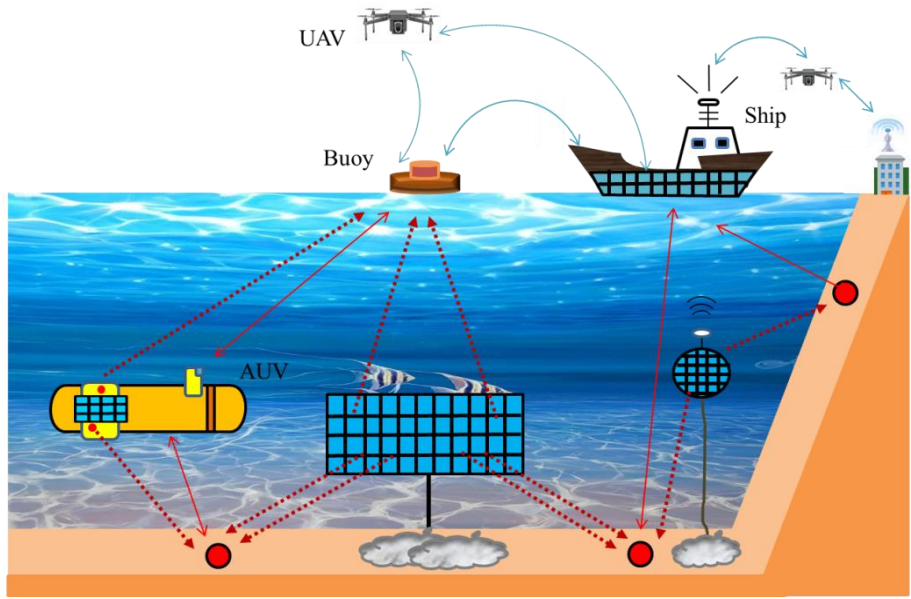

**Figure 13.** IoUT assisted by IRS-assisted UAV

**5. Open Research Challenges and Opportunities**

Though IRS-assisted UAV system seems to be a cost-efficient technology to enable smart cities including smart hospitals, smart buildings, smart factories, but still research contributions on this promising integration of these technologies is at infancy. This section elaborates some potential challenges and research opportunities which must be addressed to leverage the potentials of IRS-assisted UAVs considering both design and successful implementation.

*5.1. Physics and EM Compliant Models*

One radical challenge in existing literature on IRS is the dearth of accurate and tractable modeling which exploits reflecting metasurfaces as a feature of associated EM response. Most of the articles published on IRS assume that metasurfaces as good reflectors. However, metasurfaces must comply other features rather than just reflection. The EM behavior of metasurfaces is closely linked with different factors such as hardware design, manufacturing material, angel of incidence and reflection and polarization of impinging signal. Furthermore, the size of metasurface is also crucially significant to define its properties. Hence, EM compliant and physics modeling [12] must be considered rather than basic models with optimistic performance. Channel modeling requires proper analysis of fading effects, scattering and shadowing etc. Several parameters must be kept into consideration to design accurate models for IRS-assisted UAV channels. These parameters include IRS fabrication, geometry, number of scattering components, aerial and ground distances etc. Both IRS and UAV will put great impact to make channel modeling challenging or sophisticated. A UAV having a varying movement mobility and aerial shadowing caused by its revolving or mobility will cause wide temporal and spatial alterations. Similarly, IRS will also put its impact to define channel models due to its near-field propagation and passive reflecting behavior.

*5.2. Channel Sensing and Estimation*

IRS is comprised of multiple passive elements which are linked with a centralized controller. The usage of IRS is based on phase shifts reconfiguration of its elements to control radio channel and obtain optimal beamforming. It needs channel sensing and signal processing and channel sensing which is hard to achieve without any suitable signal processing features. In addition to this, accuracy and channel estimation delay and accuracy must be kept into consideration. Some researchers have been focusing on sophisticated signal processing algorithms to attain better accuracy and lower overhead. However, these algorithms mostly need high power consumption for computational tasks, signal processing and data transmission. Thus, an efficient and sustainable channel estimation strategy must be investigated. A possible approach is to accompany IRS elements with some low-energy active sensing devices which can sense and estimate the radio channel. Another alternative strategy is to neglect the channel estimation and rely on machine learning algorithms. For IRS-assisted UAV, accurate channel state information (CSI) has significant importance. In case of imperfect CSI acquisition, the beamforming gain will be degraded. Hence, accurate channel estimation without high power consumption and high overhead in channel training is an open research area.

*5.3. Accurate CSI Acquisition*

Generally, the feature to optimize the IRS-UAV system configuration is restricted by limited knowledge of the associated channels. Without sufficient information about the underlying propagation environment, the IRS-assisted UAV communication cannot efficiently improve the system performance. To enable the joint beamforming, the channel state information (CSI) is needed which is based on communication paths from access points (Aps) to IRS, IRS to mobile stations (MSs), APs to MSs and opposite paths. The acquisition of early and precise CSI has a major contribution in IRS-assisted UAV systems. Specially, considering indirect channels, it is hard to attain the CSI where IRS without transceiver chains is deployed. As a result the particular signal on IRS cannot be accessed. Most of the times, IRS is configured containing higher number of metasurfaces, resulting into high channel dimension and overhead on the CSI acquisition. Current studies on IRS usually suppose perfect CSI at the BS, IRS controller and users. However, it cannot be taken perfect and requires a very low overhead. Machine learning and deep learning algorithms can be used as a mitigative approach to solve this CSI challenge. In [85], authors used compressed sensing and deep learning methods to leveraging the sparsity of the channel. Despite the efficacy of ML approaches, they are not adequate for IRS-assisted UAV systems due to; (i) depending on availability of huge data to attain less generalization error; (ii) incapability to cope with the tight demands of low latency and (iii) unavailability of closed-loop optimization to deal with dynamic environments. Figure 14 shows some crucial challenges in IRS-assisted UAV channel estimation.

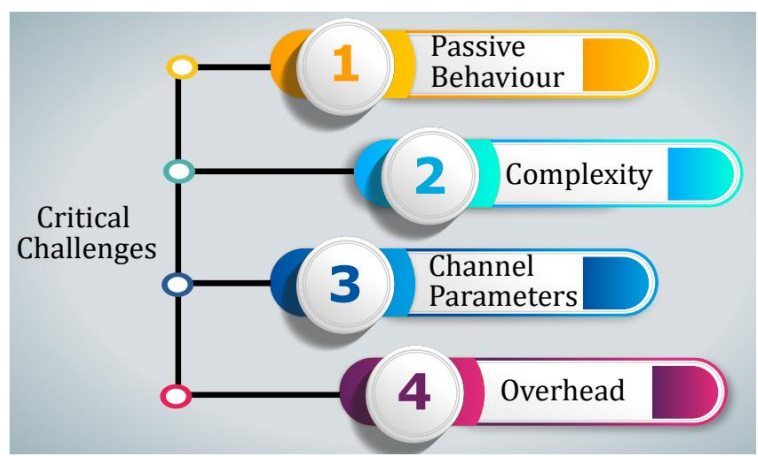

**Figure 14.** Crucial challenges in IRS channel estimation

*5.4. Controller and Overhead of IRS*

It is essential to keep account of controlling the reflecting elements of IRS. The controller is used to deliver the IRS's phase shifts. Generally, a reliable and fully synchronized controller is preferred. It can be easily implemented in static use cases or application scenario where it is easy to employ the control link. In contrast, for IRS-assisted UAV system, the control link between IRS and computing node is subject to time-dependent channel characteristics and faces shadowing or fading which affect phase shifts control in real-time cases. Additionally, IRS orientation, size and number of reflecting elements can cause a huge overhead. Thus, novel approaches are required to adjust control signaling, processing overheads to support a stable control link. To overcome these challenges, a possible approach is to consider UAV swarms with communication capabilities and distributed computing to ensure availability and reliability of control links.

*5.5. Discrete Phase Shift of IRS*

In a recent study which addresses two case studies of IRS-assisted UAV system [58], authors take continuous phase shifts of IRS during optimization of passive beamforming. Nonetheless, it appears to be an impractical approach as implementation of high-resolution reflecting elements will require hardware complexity and high cost. In such scenario, a cost-efficient approach is to employ the discrete phase shift. However, it will lead to performance degradation in contrast to continuous phase shift. Hence, it is critically important for research fraternity to propose novel strategies for discrete phase shifts at low cost, low intricacy and without comprising over system performance.

*5.6. Energy Efficiency*

Due to the absence of power amplifier, an IRS usually needs a static and limited energy supply, which is mostly ignored. On the other hand, UAV's energy conservation is of paramount importance due to deficient battery endurance and high energy consumption, which tends to be a critical bottleneck limiting the battery endurance and UAV performance. Therefore, researchers should focus on energy-efficient mechanisms with the objective to minimize the energy utility without comprising the IRS-assisted UAV communication performance.

*5.7. Reflection Efficiency*

Another major issue is to deploy IRSs in reflection-efficient technique for accurate implementations. To effective address this, the orientation and location of IRS are important features for reflection efficiency and hence required to be implemented deliberately. To overcome path loss issue at longer distances, a large size IRS with a large number of reflecting components must be adopted for passive beamforming to improve the receiver performance. However, some practical constraints such as asymmetric number of antenna elements, tunable phase geometry and duplex mode can make system highly complex. Meanwhile, it will also increase the system cost. Similarly, implementing hardware the meta-atoms, both electronically and mechanically, causes severe degradation to attain achieve adequate fast phase shifting. Thus, novel hardware designs with new materials should be investigate considering low cost and good performance.

*5.8. Environmental Factors*

In most of the previous studies, authors assume fixed location of users and stable flight of UAV. In real-time application, UAVs face inventible jittering due to vibration and airflow, tending to trivial channel estimation errors and unsteady performance. Consequently, it is hard to fully exploit the collaborative beamforming design gain. Moreover, the wind might change the UAV's trajectory and speed, leading to safety concerns and performance detriment. Furthermore, it is difficult to satisfy users QoS due to the variation in locations and dynamic environment.

*5.9. PLS against Malicious IRS-assisted UAV Attacks*

High flexibility and mobility are important in IRS-assisted UAV system to enhance the PLS and protect data leakage of legitimate users and restrict eavesdroppers. Particularly, a malicious UAV can intercept the information of legitimate UAV. In that scenario, it becomes crucial to protect the sensitive information of legitimate UAV. However, this research area is yet to be explored. Therefore, researchers should investigate advanced strategies in terms of PLS to enhance security against eavesdroppers.

*5.10. Machine Learning/Artificial Intelligence techniques for IRS-assisted UAV Communication*

It is very crucial to optimize IRS-assisted UAV communication, specifically when we deploy UAVs in a harsh or unpredictable environment. Particularly, optimizing IRS phase shift, UAV trajectory and resource allocation is hard because of non-linear models. Therefore, design strategies with efficient performance and low intricacy are required. Machine learning algorithms or artificial intelligence methods are promising strategies to design and optimize these networks. These methods offer robust and reliable tools to optimize challenging environments. In addition, data-driven and hybrid models, hybrid online and offline techniques to enhance secrecy performance might be implemented to analyze intricate networks. Howbeit, various aspects are yet required to be explored—for instance, latency, high energy consumption and large computational processing power.

**6. Conclusion**

In this study, we presented a comprehensive review of existing literature on IRS-assisted UAV communications and also investigate some new ideas. We explained both ground and airborne environments for IRS-assisted UAV communications. It can be foreseen that IRS-assisted UAV communication will expand both terrestrial and airborne wireless networks due to low cost, low energy consumption, high mobility and high reconfigurability. The basic concepts of UAV, IRS, structure, working principles and potential benefits were briefly discussed. Furthermore, this study focused on several application scenarios which can be achieved through this promising coexistence of both technologies. Additionally, we briefly explained the integration of some emerging technologies such as machine learning, free space optics, visible light communication and NOMA for breakthrough novelties of IRS-assisted UAV communication. Furthermore, we focused on critical challenges and open research issues arising from the coexistence of UAV and IRS along with solutions to overcome these challenges. Further research should be performed on real-time applications. Moreover, a real-world application for IRS-assisted UAV communication is required to evaluate the performance of the existing and future mechanisms. We believe this article will pave a way for further research contributions by providing guidance for theory, research, experimental demonstrations and real-time implementation.

**ORCID:** https://orcid.org/0000-0002-5810-4983